# Non-Stokes Drag Coefficient in Single-Particle Electrophoresis: New Insights on a Classical Problem

*Maijia Liao[1,2] (廖麦嘉), Ming-Tzo Wei[3]（魏名佐), Shixin Xu[4]（徐士鑫）,*
*H. Daniel Ou-Yang[2,3,†]（歐陽新喬）, Ping Sheng[1,†]（沈平）*

[1]*Department of Physics, Hong Kong University of Science and Technology*
*Clear Water Bay, Kowloon, Hong Kong, China*

[2]*Deaprtment of Physics,* [3]*Bioengineering, Lehigh University, Bethlehem, PA, USA*

[4]*Centre for Quantitative Analysis and Modelling (CQAM),*
*The Fields Institute, Toronto, Ontario M5T 3J1, Canada*

**ABSTRACT**

We measured the intrinsic electrophoretic drag coefficient of a single charged particle by optically trapping the particle and applying an AC electric field, and found it to be markedly different from that of the Stokes drag. The drag coefficient, along with the measured electrical force, yield a mobility-zeta potential relation that agrees with the literature. By using the measured mobility as input, numerical calculations based on the Poisson-Nernst-Planck equations, coupled to the Navier-Stokes equation, reveal an intriguing microscopic electroosmotic flow near the particle surface, with a well-defined transition between an inner flow field and an outer flow field in the vicinity of electric double layer's outer boundary. This distinctive interface delineates the surface that gives the correct drag coefficient and the effective electric charge. The consistency between experiments and theoretical predictions provides new insights into the classic electrophoresis problem, and can shed light on new applications of electrophoresis to investigate biological nanoparticles.

[†]To whom correspondence should be addressed. Emails: hdo0@lehigh.edu, sheng@ust.hk

## INTRODUCTION

When immersed in an electrolyte solution, a charged particle would be enveloped in an ionic cloud of screening counter-ions, denoted the Debye layer. Application of an external electric field to a suspension of such charged particles can result in the steady motion of the solid particulates. The physical picture underlying this phenomenon, known as electrophoresis[1-5], dates back to Smoluchowski[6] in which the crucial element is the electroosmotic fluid flow in the Debye layer. Through clever mathematical manipulations, Smoluchowski has shown rigorously that electrophoretic mobility of the charged particle, $\mu_E$, is directly proportional to the zeta potential $\zeta$ (which is directly related to the surface charge density) on the surface of the solid particle, i.e., $\bm{v}_\infty = \mu_E \bm{E}_\infty$, where $\bm{E}_\infty$ is the applied electric field, $\bm{v}_\infty$ is the electrophoretic velocity, and $\mu_E = -(\varepsilon/\eta)\zeta$, with $\eta$, $\varepsilon$ being the solution viscosity and dielectric constant, respectively. The Smoluchowski relation is accurate in the limit of $a \gg \lambda_D$, where $\lambda_D$ is the Debye length and $a$ the particle radius. A derivation of the Smoluchowski relation is given in Section S2 of Supplementary Materials (SM).

There have been extensive theoretical[6-14] and experimental[15-20] studies of a charged particle under the simultaneous effect of an electric field $\bm{E}_\infty$ and a non-electric force $\bm{F}_{ext}$. In particular, the electrophoretic drag coefficient $\gamma_E$ is of interest. This is because not only the Smoluchowski electrophoretic flow field (see Section S2 of SM) differs significantly from the Stokes flow field, but also the lack of an accurate flow field solution *inside* the Debye layer prevents an accurate account of the actual hydrodynamic drag force on the solid surface of the charged particle. A rough evaluation of drag force on the particle surface may be established through a simple scaling argument[14,21]. In the thin-Debye layer electrophoresis, the velocity gradient in the liquid is screened beyond the $\lambda_D$ scale, thus the drag force is $\sim \left[(4\pi a^2)\eta/\lambda_D\right]\bm{v}_\infty$. As $a/\lambda_D \gg 1$ in most cases, it follows that $\gamma_E \gg \gamma_S$ ($=6\pi\eta a$, the Stokes drag coefficient), with the ratio $\gamma_E/\gamma_S$ reaching as high as 100 to 200 in some cases. However, our accurate measurements of $\gamma_E$, described below, does not come anywhere close to such order of magnitude deviation from the Stokes drag. This large discrepency necessitates an understanding of the physical picture underlying the electrophoretic drag.

Based on the linearization of coupled electrohydrodynamic equations, plus the superposition of external non-electric force $\boldsymbol{F}_{ext}$ and the electrophoresis problem[1,21], it is found that a charged particle behaves similarly in an electric field as in a hydrodynamic flow[22]. From force balance one obtains:

$$\boldsymbol{F}_{ext} - \gamma_s(\boldsymbol{v} - \mu_E \boldsymbol{E}_\infty) = 0 \quad , \tag{1}$$

where $\boldsymbol{v}$ denotes the solid particle velocity under the combined electric and non-electric (mechanical) forces. Equation (1) suggests that the mechanical force needed to stall an electrophoretic motion is a Stokes-like drag force.

In anticipation of subsequent developments, we make two remarks in relation to Eq. (1). The first is that Eq. (1) can be expressed in terms of two equations by the velocity decomposition $\boldsymbol{v} = \boldsymbol{v}_\infty + \Delta \boldsymbol{v}$, where the first equation, $\boldsymbol{v}_\infty = \mu_E \boldsymbol{E}_\infty$, is the definition of the electrophoretic velocity. The second equation is then $\boldsymbol{F}_{ext} - \gamma_s \Delta \boldsymbol{v} = 0$. Such a decomposition is useful in demonstrating that the electrophoretic drag coefficient $\gamma_E$ is not necessarily identical to $\gamma_S$. This can be easily seen by multiplying the velocity equation on both sides by $\gamma_E$: $\boldsymbol{F}_E = \gamma_E \boldsymbol{v}_\infty = \gamma_E \mu_E \boldsymbol{E}_\infty = Q_{eff} \boldsymbol{E}_\infty$. Here $\gamma_E$ is simply defined to be the coefficient of proportionality between velocity and hydrodynamic viscous force, in this case both resulting from an applied electric field; and $Q_{eff} = \gamma_E \mu_E$ is introduced to distinguish it from the surface charge, since $Q_{eff}$ is known to be much smaller than the surface charge[14,16] and represents, in the context of force balance, the solid particle's coupling to the applied electric field.

The second remark is related to our experimental approach of using an optical trap to hold a single charged particle in a harmonic potential and applying an AC electric field to induce periodic oscillations of the particle. The accurately measured quantities are then the amplitude of particle's periodic motion and its phase difference with the applied AC electric field. Owing to the AC nature of the applied electric force, there are inevitably the in-phase and out-of-phase components of the force relative to the particle velocity. For Eq. (1), i.e., $\boldsymbol{F}_{ext} - \gamma_s \Delta \boldsymbol{v} = 0$, the applied $\boldsymbol{F}_{ext}$ is necessarily the in-phase component of the force; it describes that the force needed to alter, or stall, an intrinsic electrophoretic motion is Stokes by nature as argued by Long et al.[21] and thus, the equation itself has nothing to do with electrophoresis. Our simulation results (see below) supports the correctness of Long's argument, but contradicts the non-Stokes prediction of the in-phase external force by Lizana et al.[14]. We will show below in the subsequent section that in our experiment the *optical trapping force is always out-of-phase relative to the phase of the velocity*, whereas the applied electric force has both an in-phase component and an out-of-phase component. The in-phase component of the electric force drives the particle velocity and the out-of-phase component automatically counter-balances the out-of-phase optical trapping force. These facts account for our approach's ability to measure the electrophoretic drag force and its drag coefficient.

The drag coefficient is always related to the hydrodynamic drag force exerted on a surface. As Smoluchowski has successfully linked $\mu_E$ to the surface zeta potential, and measured drag force is given by $\gamma_E \boldsymbol{v}_\infty = \gamma_E \mu_E \boldsymbol{E}_\infty = Q_{eff} \boldsymbol{E}_\infty$, achieving force balance between the electrical force and drag force at the solid surface (by equating $Q_{eff}$ to $Q_S$) would seem to be the most convenient choice. However, the fact that $Q_{eff}$ is known to be much smaller than $Q_S$ [14,16], which is also confirmed by our measurements as seen below, indicates that there must be another surface, away from the solid surface[24-26], on which the drag force and electrical force attain force balance. Two questions naturally arise: (1) How does such a surface emerge consistently from the relevant mathematical equations governing the electrophoresis, and (2) Can one measure an electrophoretic drag coefficient $\gamma_E$ that is consistent with the theory prediction on such a surface?

To address these two questions, we set out to measure $\gamma_E$ and $Q_{eff}$ directly by experiments, and to obtain mathematically accurate simulations of the electrophoretic flow field. From experimental measurements the mobility was obtained as $\mu_E = Q_{eff} / \gamma_E$. It is shown that while $\gamma_E / \gamma_S$ varies only from 1 to 3 over a range of salt concentrations, the magnitude of $Q_{eff}$ is smaller than that of $Q_S$ by many orders of magnitude. The resulting $\mu_E$, however, agrees well with the literature values on similar micro-spheres, as well as that obtained by DC measurements in our experiments[26]. To provide a microscopic explanation of our experimental data, we have carried out simulation of the electrophoresis effect by numerically solving the Poisson-Nernst-Planck (PNP) equations coupled with the Navier-Stokes (NS) equation, implemented with the appropriate boundary conditions. Using only the experimentally measured $\mu_E$ as input, the most significant outcome is that the measured $\gamma_E$ and $Q_{eff}$ are obtained at a uniquely distinctive interface between the inner and outer flow fields that is at a distance a few $\lambda_D$'s away from the solid surface. While the electrophoretic flow in the outer flow region follows the scaling of the Smoluchowski solution, i.e., $r^{-3}$, our numerical solution reveals an inner flow field *that is carried along by the solid particle*, with a highly nonlinear electro-hydrodynamic flow behavior. The interface between the inner and outer flow fields acts as the reference surface, or slip surface/plane, at which both the magnitude and trend (e.g., with respect to salt concentration) of the drag coefficient and effective charge can be accounted for, and from which the deviations can be further explored.

In what follows, we first present the experimental results, followed by simulations and discussion of the physical picture that emerges. We end by underscoring the closure between theory and experiment.

**EXPERIMENTAL RESULTS**

**Optical trap and AC electric field**

We apply an AC electric field (20-100 Hz) to a single spherical particle held by a calibrated optical trap as shown in Fig. 1(a). At such frequencies the period of the applied field was much smaller than the time required to screen the electrodes (>1 s for a separation between the electrodes=1 cm), and much larger than the relaxation time of the electrical double layer[27] (<1 μs). The relevant equation of motion is given by:

$$m\ddot{x}(t) = -\gamma \dot{x}(t) + E_\infty Q_{eff} - k_{trap} x(t). \qquad (2)$$

Here, $m$ is the mass of the particle, $x$ is the displacement of the particle from the center of the harmonic optical trap, and $\gamma$ is the drag coefficient. Here the drag coefficient is purposely written without a subscript, since in Eq. (2) there are two forces—optical trap force and electric force—acting on the particle and hence the specification of which drag coefficient applies would depend on the situation to be analyzed below. The Brownian noise term is excluded in the above equation because its contributions will be filtered out by the phase-sensitive, lock-in detection technique employed in our experiments. Here $Q_{eff}$ is the effective charge that provides particle's electrical coupling to the applied electric field. It should be noted that the deformation of the counterion cloud is very small, so that the resulting electric "retardation" force can be neglected (*28*). Our simulations, based on the full numerical solution of the relevant governing equations, have confirmed the negligible effect of retardation.

By approximating the left-hand side of Eq. (1) to be zero, which is accurate considering the fact that $m\omega^2$ is 5 orders of magnitude lower than $k_{trap}$, we have:

$$k_{trap} x(t) = -\gamma \dot{x}(t) + E_\infty Q_{eff}, \qquad (3)$$

where $E_\infty = E_\infty^{(0)} \exp(-i\omega t)$. We write the displacement of the electrophoretic particle as $x(t) = D(\omega)e^{-i[\omega t - \delta(\omega)]}$, with the displacement amplitude denoted by $D(\omega)$ and the phase $\delta(\omega)$ defined relative to the applied AC electric field. In our experiments, the phase shift was measured with a lock-in amplifier with an accuracy of a couple of degrees. This is in contrast to the phase shift measurements in a previous paper[15] that has errors in the range of +/− 11 degrees. Through the precise measurement of the phase shift, our method enabled the extract of drag coefficient value in an accurate and robust manner.
.

**An analysis of the AC experimental approach**

Substituting particle displacement expression for $x(t)$ and the associated displacement velocity $\dot{x}(t)$ back into Eq. (3), and taking the real part of every term, we get

$$k_{trap}D(\omega)\cos[\omega t - \delta(\omega)] = \gamma\omega D(\omega)\sin[\omega t - \delta(\omega)] + E_\infty^{(0)}Q_{eff}\cos(\omega t). \quad (4)$$

Let us define a $t_0$ such that $\omega t_0 = \delta(\omega) + \dfrac{\pi}{2}$, and let $\omega t = \omega(t_0 + t') = \omega t_0 + \omega t'$. Equation (4) can be re-written in an illuminating form as

$$k_{trap}D(\omega)\sin(\omega t') = -\gamma\omega D(\omega)\cos(\omega t') + E_\infty^{(0)}Q_{eff}\sin[\omega t' + \delta(\omega)]. \quad (5)$$

At $t'=0$, the left-hand side vanishes, i.e., optical trap force is zero, and we only have the right-hand side, from which we obtain

$$\gamma_E \omega D(\omega) = E_\infty^{(0)}Q_{eff}\sin[\delta(\omega)]. \quad (6)$$

We label the drag coefficient as $\gamma_E$ because there is only the electric field force present in Eq. (6). Now let us expand $\sin[\omega t' + \delta(\omega)] = \sin(\omega t')\cos[\delta(\omega)] + \cos(\omega t')\sin[\delta(\omega)]$ in Eq. (5). Then Eq. (5) can be re-organized in a physically clear manner as

$$[k_{trap}D(\omega) - E_\infty^{(0)}Q_{eff}\cos\delta(\omega)]\sin(\omega t') = [-\gamma_E \omega D(\omega) + E_\infty^{(0)}Q_{eff}\sin\delta(\omega)]\cos(\omega t'). \quad (7)$$

In Eq. (7), the right- and left-hand sides represent the in- and out-of-phase components, respectively. There is no approximation made in Eq. (7), thus it has to be valid for all values of $t'$. The only way Eq. (7) can be true is that both sides must separately be zero, since the time variations on the two sides are orthogonal to each other.

Physically, Eq. (7) states that the electrical force comprises two components. One component, $E_\infty^{(0)}Q_{eff}\sin\delta(\omega)\cos(\omega t')$, gives rise to the time-varying electrophoretic velocity $\omega D(\omega)\cos(\omega t')$ with a constant drag coefficient $\gamma_E$. The other component, $E_\infty^{(0)}Q_{eff}\cos\delta(\omega)\sin(\omega t')$, counter-balances the optical trap force $k_{trap}D(\omega)\sin(\omega t')$. This optical trap response force is noted to be *out of phase* with the electrophoretic velocity, in contrast to the external force in Long's equation (Eq. (1)), which is in-phase with the electrophoretic velocity.

Equation (7) essentially expresses the fact that 0=0, from which we obtain two independent equations:

$$k_{trap}D(\omega) = E_\infty^{(0)}Q_{eff}\cos\delta(\omega), \quad (8a)$$

$$\gamma_E \omega D(\omega) = E_\infty^{(0)}Q_{eff}\sin\delta(\omega). \quad (8b)$$

Equation (8b) is identical to Eq. (6). And if we divide (8b) by (8a), we obtain the consistency condition imposed by the force balance of the in- and out-of-phase components of the AC experiment:

$$\gamma_E = \frac{k_{trap} \tan \delta(\omega)}{\omega} . \qquad (9a)$$

Equation (8a) can be re-written for $Q_{eff}$ in terms of the physically measured quantities as

$$Q_{eff} = \frac{k_{trap} D(\omega)}{E_\infty^{(0)} \cos \delta(\omega)} . \qquad (9b)$$

From Eqs. (9a) and (9b) the mobility is obtained as

$$\mu_E = Q_{eff} / \gamma_E = \frac{\omega [D(\omega) / E_\infty^{(0)}]}{\sin[\delta(\omega)]} . \qquad (9c)$$

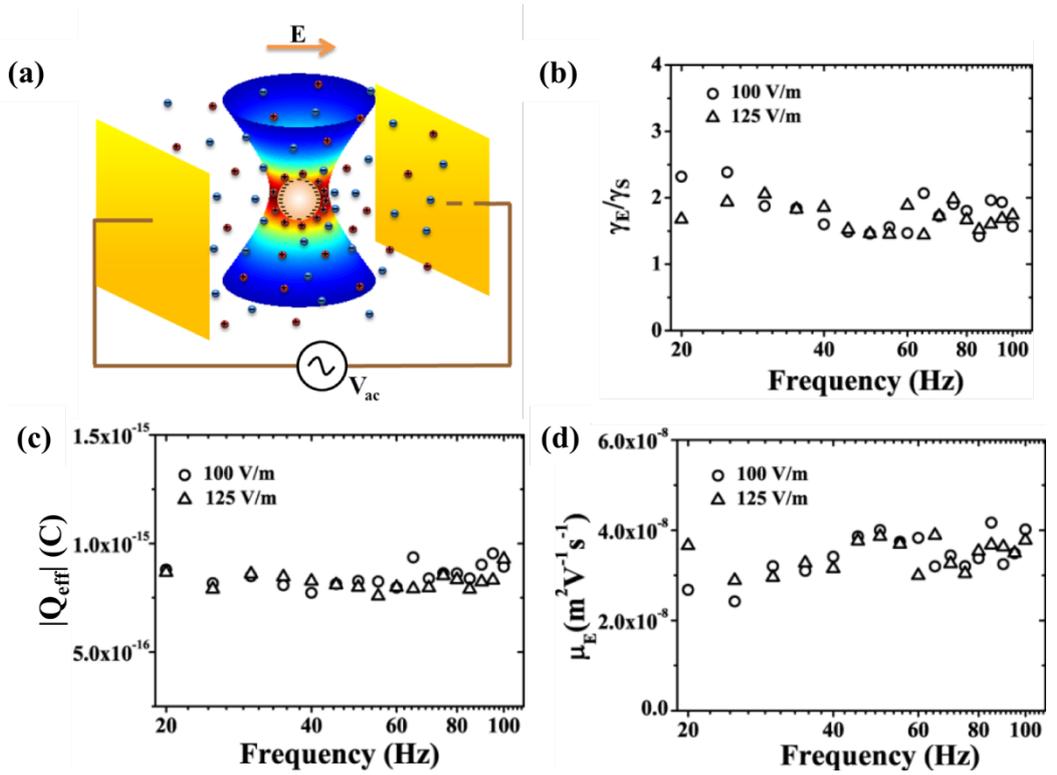

**Fig. 1. Electrophoretic measurements using the optical tweezers.** (*a*) Schematic illustration of a charged colloidal particle held by an optical tweezer and driven by an oscillating electric field. The electric field direction is indicated for a particular instant of time. The measured $\gamma_E/\gamma_S$ (**b**), magnitude of $Q_{eff}$ (**c**), and $\mu_E$ (**d**) plotted as a function of applied electric field frequency for a 0.75 μm radius polystyrene particle, held by a fixed optical trap in DI water ($\lambda_D$=96.1nm) under two different electric field strengths. It is seen that all measured quantities are relatively independent of the frequency, indicating the quasi-static nature of the experiment.

Not limited by the constraint imposed by typical DC measurements that any external non-electrical force would alter the intrinsic electrophoretic motion and yield a Stokes-like drag, the most essential point of the above analysis is that the external force imposed by the optical trap is always 90 degrees out of phase with the electrophoretic velocity. Through this feature of the AC electric field-driven particle in an optical harmonic trap, we can measure electrophoresis drag coefficient $\gamma_E$ accurately without the influences of an in-phase, non-electrical external force.

**Measured drag coefficient, effective charge and mobility**

Using the measured values $D(\omega)$ and $\delta(\omega)$ in Eqs. 9(a) and 9(b), respectively, we have evaluated $\gamma_E$ and $Q_{eff,}$ which are plotted in Figs. 1(b) and (c) as a function of applied electric field frequency for a 0.75 μm radius polystyrene particle, held by a fixed optical trap in de-ionized (DI) water ($\lambda_D$=96.1nm) under two different electric field strengths. In Fig. 1(d) we show the measured values of $\mu_E$ plotted as a function of applied electric field frequency, also for two different electric field strengths. It is seen that the measured $\gamma_E$, $Q_{eff}$ and $\mu_E$ were relatively independent of the frequency and the field strength, indicating the quasi-static nature of our measurements; we take as measured values the averages over the measured frequency range.

It should be noted here that although the electrophoretic mobility determined from Eq. (9c) was determined from the AC measurements, we have also measured it using the traditional approach by using a DC field. Experimentally, this was done by turning off the optical trap and measuring the speed of the same particle in the presence of a DC electric field. The mobility determined by the AC measurements agreed with that by the DC measurements, as expected. It should be noted that the more precise phase shift values in our measurements clearly show the phoretic drag to be non-Stokes, while simultaneously the electrophoretic mobilities obtained by our AC and DC measurements agree within the experimental error. This is in contrast to the earlier experiment by Semenov et al.[15], in which the phase shift measurement had a much larger error bar. Moreover, by assuming the drag coefficient to be Stokes, Semenov's study yielded different mobilities between their AC and DC experiments.

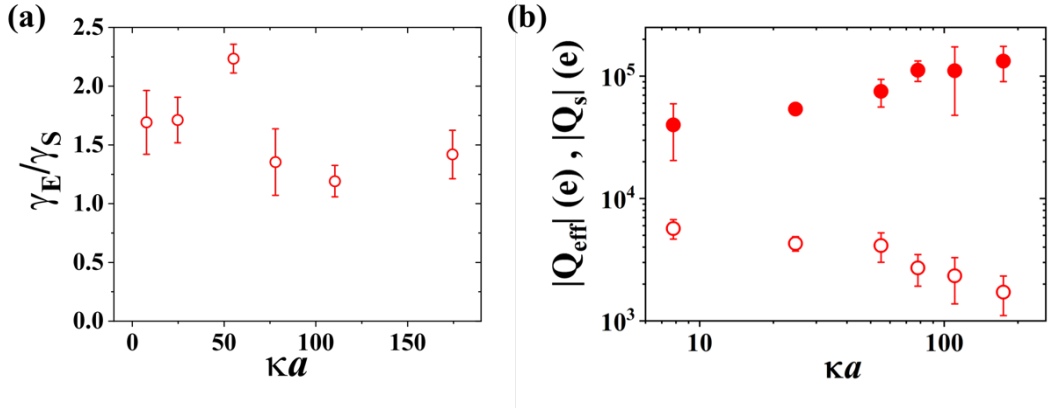

**Fig. 2 .Measured drag coefficient and effective charge plotted as a function of salt concentration, expressed here as** $\kappa a$ **, where** $\kappa = 1/\lambda_D$ **. (a)** Open symbols denote the values of measured $\gamma_E/\gamma_s$ plotted as a function of $\kappa a$. Here $\gamma_s$ denotes the calculated Stokes drag coefficient, and $\gamma_E$ is obtained by using measured $\delta$ in Eq. (9). Error bars on the open circles arise from the errors in amplitude and phase measurements. **(b)** Measured effective charge magnitude $|Q_{eff}|$ is plotted as a function of $\kappa a$ with open symbols. Filled symbols indicate the surface charge magnitude $|Q_s|$ obtained from simulations with measured mobility $\mu_E$ as the only input from the experiment. Error bars are from the mobility measurements. All results are for PS spheres with $a$=0.75 μm.

In Fig. 2(*a*) the measured $\gamma_E/\gamma_s$ is plotted as a function of κ*a* with open symbols, where $\kappa = 1/\lambda_D$. Non-Stokes drag coefficients values are clearly observed. As κ*a* increases towards 110, however, $\gamma_E$ gradually approaches $\gamma_S$ which is in direct contrast to the $4\pi\eta a^2/\lambda_D$ value of the scaling argument in the thin Debye layer limit[14,21].

Values of the measured magnitude of the effective charge $|Q_{eff}|$ are plotted with open symbols as a function of ionic strength in Fig. 2(b). They are in the range of 1700 to 4800 electronic charges, which are one to two orders of magnitude smaller than the generally accepted values deduced from known PS surface charge density[26,29,30] that is in the range of ~ $-6\times10^4$ *e/μm²*. For a 0.75 μm radius PS spheres, one obtains $|Q_s| \sim 1.9\times10^5$ *e*, where *e* denotes the magnitude of the electronic charge. In Fig. 2(b), the magnitude of the surface charge $|Q_s|$ obtained from our numerical simulations, obtained with the experimentally measured mobility values as inputs, are shown by solid symbols; these values are noted to be in the expected range of $|Q_s|$ for PS surfaces, thus further confirming the huge differences with the magnitude of the effective charge. Also, as $Q_s$ is always directly related to the mobility, the resulting zeta potential/mobility relation is compared to the literature in Section S5 of SM. Excellent agreement is obtained.

**SIMULATIONS**

**Equations and boundary conditions**

Our starting point is the coupled incompressible NS equation and the PNP equations for an electrolyte that is symmetric between the positive and negative ions. The equations are as follows.

$$\frac{\partial n, p}{\partial t} = -\nabla \cdot \boldsymbol{J}_{p,n}, \tag{10a}$$

$$\boldsymbol{J}_n = -D_{ion}\left(\nabla n(\boldsymbol{r}) - \frac{e}{k_B T} n(\boldsymbol{r})\nabla \psi(\boldsymbol{r})\right) + n(\boldsymbol{r})\boldsymbol{u}(\boldsymbol{r}), \tag{10b}$$

$$\boldsymbol{J}_p = -D_{ion}\left(\nabla p(\boldsymbol{r}) + \frac{e}{k_B T} p(\boldsymbol{r})\nabla \psi(\boldsymbol{r})\right) + p(\boldsymbol{r})\boldsymbol{u}(\boldsymbol{r}), \tag{10c}$$

$$\nabla^2 \psi = -\frac{e(p-n)}{\varepsilon_f \varepsilon_0} \tag{10d}$$

$$\rho_\ell \frac{\partial \boldsymbol{u}}{\partial t} = \eta \nabla^2 \boldsymbol{u} - e(p-n)\nabla \psi - \nabla P, \tag{10e}$$

$$\nabla \cdot \boldsymbol{u} = 0, \tag{10f}$$

$$\boldsymbol{F}_E = \int \boldsymbol{T}^E \cdot \hat{\boldsymbol{n}} d\Gamma = \int \varepsilon_0 \varepsilon_f \left[\boldsymbol{E}\boldsymbol{E} - \frac{1}{2}(\boldsymbol{E} \cdot \boldsymbol{E})\boldsymbol{I}\right] \cdot \hat{\boldsymbol{n}} d\Gamma, \tag{10g}$$

$$\boldsymbol{F}_H = \int \boldsymbol{T}^H \cdot \hat{\boldsymbol{n}} d\Gamma = \int \left[-P\boldsymbol{I} + \eta\left(\nabla \boldsymbol{u} + \nabla \boldsymbol{u}^T\right)\right] \cdot \hat{\boldsymbol{n}} d\Gamma, \tag{10h}$$

$$m\frac{d\boldsymbol{v}}{dt} = \boldsymbol{F}_E + \boldsymbol{F}_H. \tag{10i}$$

Here $\psi$ stands for electrostatic potential, $p$ ($n$) stands for the positive (negative) ion density, $\boldsymbol{J}_{p(n)}$ stands for ionic current density, with the diffusive, electric convective, and flow convective components, $\eta = 1 \times 10^{-3}$ Pa.s is the liquid viscosity, taken to be that of water, $\rho_\ell$ is the liquid density of water, the dielectric constant of water, $\varepsilon_0 \varepsilon_f$, is taken to be $\varepsilon_f = 80$, with $\varepsilon_0 = 8.85 \times 10^{-12}$ F/m, $\boldsymbol{u}$ is the liquid velocity, and $P$ denotes pressure. The valence of all ions is taken to be one. Here we use the value of $D_{ion} = 9.32 \times 10^{-9} m^2/s$ [31]. The boundary conditions for Eqs. 10(a)-(f) are $\psi(r) \to -\boldsymbol{E}_\infty \cdot \boldsymbol{r}$ as $r \to \infty$, $\hat{\boldsymbol{n}} \cdot \boldsymbol{J}_{n/p}|_{r=a} = 0$, where $\hat{\boldsymbol{n}}$ stands for the unit outward surface normal, and $\boldsymbol{E}_\infty$ is the applied electric field. Here infinity denotes the simulation domain boundary. In Eqs. (10g) and (10h), $\boldsymbol{E} = -\nabla \psi$ and $\Gamma$ stands for the particle surface, $\boldsymbol{F}_{E(H)}$ stands for the electric (hydrodynamic viscous) force, and $\mathbf{T}^{E(H)}$ denotes the electric (hydrodynamic viscous) stress tensor. Equation (10i) is the Newton's equation for particle motion.

Static simulations were performed in the co-moving particle frame with time derivatives in Eqs. (10a)-(10i) set to zero. Constant surface charge boundary condition was used for the boundary condition on the solid particle surface, with its value adjusted according to the criterion that $\boldsymbol{F}_E + \boldsymbol{F}_H = 0$ on the sphere surface at steady state, subject to the velocity constraint $\boldsymbol{u}(\boldsymbol{r})|_{r\to\infty} = -\mu_E \boldsymbol{E}_\infty$ at the simulation domain boundary, where $\mu_E = Q_{eff}/\gamma_E$ is the measured mobility. The ion densities at simulation domain boundary are given by $n(\boldsymbol{r})|_{r=\infty} = n^\infty$, $p(\boldsymbol{r})|_{r=\infty} = p^\infty$, with equality between the two as required by overall charge neutrality. The boundary condition at the fluid particle interface is non-slip. The pressure at the simulation boundary is set as constant. For dynamic simulations as shown below, the initial velocity of the particle is set to be zero. A time varying electric field is applied to drive the particle motion. More details on static and dynamic simulations are presented in Section S5 of SM.

**Physical picture and discussion**

We would like to describe the salient features of the electrophoretic flow pattern, and to contrast them with those of the Stokes flow field. For a spherical particle acted on by an external force along the direction of unit vector $\hat{\boldsymbol{F}}$, the Stokes flow field in the lab frame[32] is given by:

$$\boldsymbol{u}(\boldsymbol{r}) \cdot \hat{\boldsymbol{F}} = u_0 \left[ \frac{1}{4}\left(\frac{a}{r}\right)^3 \left(1 - 3\cos^2\theta\right) + \frac{3}{4}\left(\frac{a}{r}\right)\left(1 + \cos^2\theta\right) \right], \quad (11a)$$

where $u_0$ denotes the particle velocity, $r$ denotes the radial coordinate, and $\theta$ the polar angle defined relative to $\hat{\boldsymbol{F}}$. In contrast, the Smoluchowski fluid velocity field for particle electrophoresis in the lab frame can be written as (see Section S2 of SM):

$$\boldsymbol{u}(\boldsymbol{r})\cdot\hat{\boldsymbol{E}}_{\infty} = -\frac{\varepsilon\zeta}{\eta}\boldsymbol{E}_{\infty}\cdot[\frac{1}{2}\left(\frac{a}{r}\right)^3\left(\vec{I}-3\hat{r}\hat{r}\right)]\cdot\hat{\boldsymbol{E}}_{\infty} = -\frac{\varepsilon\zeta}{\eta}\left[\frac{1}{2}\left(\frac{a}{r}\right)^3\left(1-3\cos^2\theta\right)\right]E_{\infty}, \qquad (11b)$$

where $\hat{\boldsymbol{E}}_{\infty}$ denotes the unit vector along the direction of the externally applied electric field, taken as the $z$ direction. It is seen that the angular part of Eq. (11b) represents the second order Legendre polynomial $P_2(\cos\theta)$. We use the angular profile of the Smoluchowski flow field to project the simulated velocity field, i.e., $U(r) = \int_{-1}^{1}\boldsymbol{u}_S\cdot\hat{\boldsymbol{E}}_{\infty}P_2(\cos\theta)d\cos\theta$, in which $\boldsymbol{u}_S$ is the simulated velocity field. The plotted blue curve in Fig. 3(a) is $U(r)/v_{\infty}$. The most significant feature to be noted in Fig. 3(a) is that the $1/r^3$ behavior of the radial asymptotic flow field predicted by the Smoluchowski is indeed very well re-produced by the projected result. However, in contrast to the Smoluchowski solution, the $1/r^3$ behavior of our simulated flow field does not extend to the surface of the solid sphere. Instead, there is a sharp drop in the projected velocity field at a small distance away from the solid surface that is on the order of a few Debye lengths. We denote the peak in the projected velocity field in Fig. 3(a) to be the interface between the inner and outer flow fields. The fact that there is such an inner flow field that differs from the (Smoluchowski) outer flow field should not be surprising, since the inner region is dominated by the existence of the Debye layer whose static differential equation (derivable from the PNP equations), the Poisson-Boltzmann equation[31,33], is already highly nonlinear. Hence mathematically there should be a qualitatively distinguishable transition as the radial coordinate $r$ approaches to within a few Debye lengths.

In Fig. 3(b) we show the full simulated electrophoretic flow field in the laboratory frame. It is noted that the streamlines of the electrophoretic flow pattern in Fig. 3(b) exhibits a belt of stationary vortices around the equatorial plane of the particle, with the electric field direction defining the north and south poles. The interface between the inner and outer flow fields is delineated here by the white dashed line. In contrast, in Fig. 3(c) we show the Stokes flow field with the same-sized particle moving at the same velocity as in (b). There are no vortices in the flow field. The difference in the magnitude of the far field velocity is also apparent, as indicated by color, owing to the much slower $1/r$ velocity decay of the Stokes flow. Since the Reynolds number is negligibly small in electrophoretic flow, here the existence of vortices seen in 3(b) is due to the large net charge and the associated strong local electric field in the Debye layer. It is known mathematically that the conditions for the existence of vortices are (1) the existence of a source term for vorticity ($\omega = \nabla \times v$) as derived from the NS equation, and (2) velocity reversal since the velocity on two sides of a vortex must be opposite. The latter condition is guaranteed in the laboratory frame of the electrophoretic flow close to the solid boundary and in the vicinity of the equatorial plane, because the electroosmotic flow induced by the net (positive) charge within the Debye layer is opposite to that of the solid particle. Hence both conditions are satisfied for the electrophoretic motion in the laboratory frame, but not in the co-moving frame due to the lack of velocity reversal. Hence vortices are not apparent for the electrophoretic motion in the co-moving frame. For the Stokes flow none of the two conditions is satisfied in the laboratory frame. Mathematical details for the non-inertial generation of vorticity, as in the present case, are shown in Section S6 of SM.

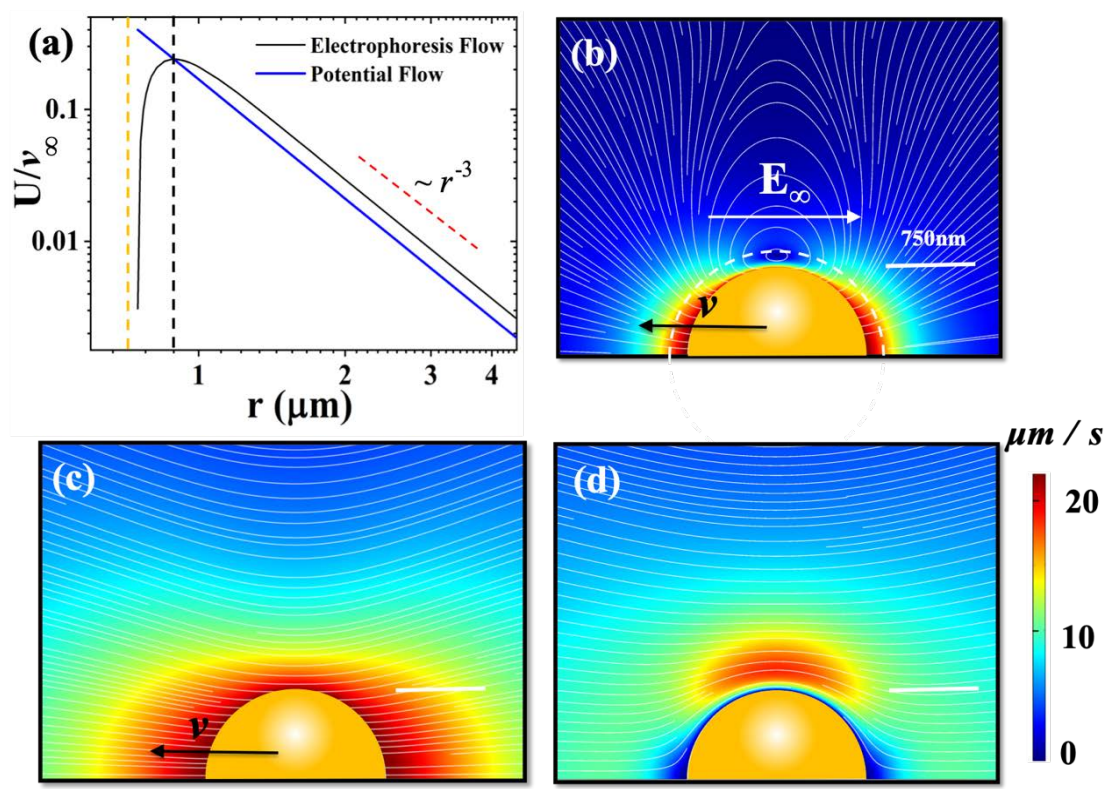

**Fig. 3 Projected velocity and streamlines in the laboratory frame for same-sized particles under stationery condition (force balance with no acceleration).** The white bars in the figures denote the length scale of 750 nm. **(a)** $P_2$ projection of the electrophoretic velocity component along the electric field direction normalized by electrophoresis velocity $v_\infty = \mu_E E$ shows very good far-field $1/r^3$ behavior, indicated by the red dashed line. Yellow dashed line shows the position of particle surface. Blue line shows the $P_2$ projection of the Smoluchowski potential flow velocity field expressed by Eq. (11b). It is clear that the projected $1/r^3$ behavior of the simulated velocity field stops at a small distance away from the solid surface. The peak position of the project result is delineated by the black dashed line, corresponding to the white dashed curve in (b). **(b)** Streamlines for the negatively charged PS sphere with $a=0.75$ μm, (with $\lambda_D=96.1$ nm, $\kappa a=7.8$, $\sigma=7000e/\mu m^2$) plotted in the lab frame. Here the colors indicate the magnitude of the velocity; the interface between the inner and outer flow fields (the reference surface) is denoted by white dashed curve. **(c)** Streamlines for a particle of the same radius as (a), acted on by a constant external body force to moves at same velocity as $\mu_E E_\infty$. This represents the Stokes flow field in the lab frame. The contrast with the electrophoretic flow field shown in (b) is clear. **(d)** Streamline for the flow field obtained from a charged particle under the same electric field strength as in (b) but fixed by a reverse external (non-electrical) force acting on the particle. This flow field can also be obtained approximately by the superposition of the electrophoretic flow field in (b) and the Stokes flow field in (c) with a reversed velocity. It should be noted that the external (non-electrical) force required to immobilize the particle is not the same as the electrical force, thereby explaining the difference in the drag coefficients. The difference in the two forces accounts for the non-zero flow field close to the particle.

The physical manifestations of the inner flow field, as afforded by the full numerical solution of the relevant governing equations, represents a realization of the Smoluchowski picture in which the electroosmotic flow near the particle surface is the dominant mechanism of electrophoresis. In the original Smoluchowski solution (see Section S2 of SM) the inner flow field region was compressed to infinitesimal thickness for the sake of mathematical simplicity. Here we restore the inner flow field region with all its rich electro-hydrodynamic flow behavior.

To further accentuate the difference between the electrophoretic and Stokes flows and their respective drag coefficients, in Fig. 3(d) we show the simulated flow field of a charged particle under the same applied electric field strength as in 3(b) but fixed by an external non-electrical body force acting on the particle. This flow field can also be obtained approximately by the superposition of the electrophoretic flow field in 3(b) and the Stokes flow field in 3(c) with a reversed velocity. The fact—that the flow field in 3(d) is not identically zero—emphasizes our point that the external non-electrical body force required to immobilize the particle is *not* the same as the electrical force. It also provides another, and perhaps more direct, explanation for why the drag coefficient in Eq. (1) is not the same as the electrophoretic drag coefficient. The difference between the Stokes vs. the electrophoretic drag forces accounts for the non-zero flow field close to the particle in 3(d); even though the net force is evaluated to be zero on the solid surface, as required by force balance on an immobile particle.

**CLOSURE BETWEEN THEORY AND EXPERIMENT**

**Drag coefficient under different $\kappa a$ values**

In Fig. 4 we summarize the overall results of this work. In Fig. 4(a) the ratio between the measured drag coefficients and the Stokes drag coefficient is plotted as open squares onto the solid simulation curves with different $\kappa a$ values, each evaluated on a surface at distance $r$ outside the solid particle's surface. The vertical dashed line indicates the inner/outer interface position, whose intersections with the different curves yield the drag coefficients at that interface. The measured results cluster around, and are close to, the values at the inner/outer interface. Their values are significantly smaller than the simulated values at the solid surface. The solid squares are calculated from the scaling argument $\left[(4\pi a^2)\eta/\lambda_D\right]/\gamma_S$ [14,21]. It is not surprising that these values at the solid surface are much larger than $\gamma_S$, since $a/\lambda_D$ is in the range of 7-200 in our experimental measurements.

In Fig. 4(b) we plot the position of the inner/outer flow field interface, in terms of its dimensionless distance $\kappa\Delta$ away from the particle surface, as a function of the salt concentration. Solid symbols indicate the inner/outer interface position as determined by the peak values in $U/v_\infty$ as shown in Fig. 3(a). The open symbols represent the values of $\Delta$ determined by using measured drag coefficients as input to find the spherical surfaces on which the calculated drag forces exactly matches the measured values. Both the magnitudes and the trend of the empty symbols track the inner/outer interface position very well. Note that $\Delta$ is independent of surface charge density under a given particle radius.

The fact that the experimentally measured drag coefficient yields a surface position whose salt concentration dependence is identical to that of the inner/outer interface, ties the latter rather uniquely to its role as the reference surface for the electrophoretic drag and effective charge.

Below we further reinforce this point by linking this inner/outer interface to the theoretical framework of O'Brien and White[9], and Long et al.[21]

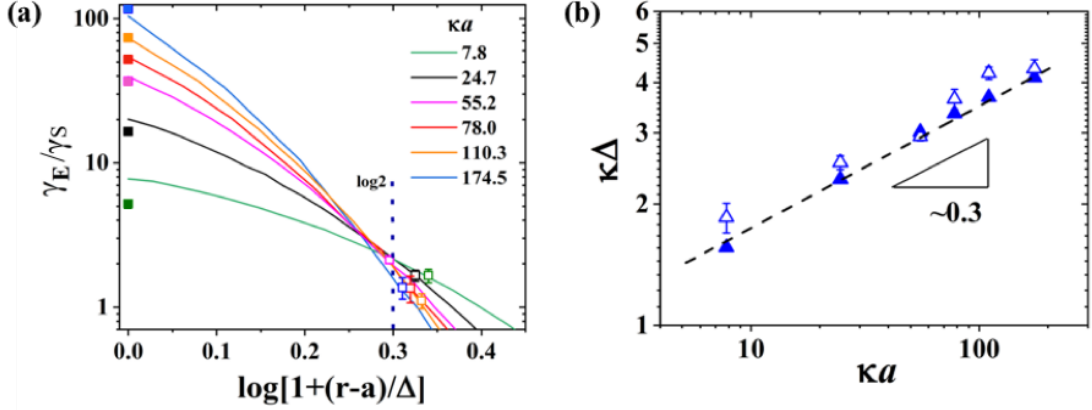

**Fig. 4 (a)** Drag coefficients calculated at various spherical surfaces at distance $r$ outside the solid particle are represented by solid curves. Measured drag coefficients are plotted as open symbols. The inner/outer interface position is denoted as the dashed vertical blue line at log2. The solid symbols at $r=a$ indicate the values of $\left[(4\pi a^2)\eta / \lambda_D\right]/\gamma_S$ which is an estimation for electrophoresis drag coefficient in previous literature[14,21]. **(b)** Values of $\kappa\Delta$, which indicates the reference surface position $\Delta$ normalized by $\lambda_D$, are plotted as a function of $\kappa a$. Solid symbols indicate the inner/outer interface position as determined by the peak values in $U/v_\infty$ as shown in Fig. 3(a). Open symbols indicate $\kappa\Delta$ values as determined by using the measured drag coefficients as the inputs to find the resulting interface position on which the calculated drag coefficient exactly yields the experimental value. The proximity and the trend give no doubt that, as far as drag coefficient is concerned, the inner/outer interface is the relevant reference surface at which the drag force is being acted on. The dashed line is a guide to the eye; it follows the relation $\Delta=0.87a^{0.3}\lambda_D^{0.7}$.

**Dynamic simulation and comparison with measured data**

To further substantiate the physical picture that the measured hydrodynamic drag coefficient should be evaluated by using the inner/outer interface (dashed white curve in Fig. 3(b)) as the reference (slip) surface/plane, we have carried out dynamic simulations by using the moving mesh approach. Details of the simulations are given in Section S6 of SM. In Fig. 5 the open circles are the results of the dynamic simulation plotted as a function of time. The applied electric field is a step function with a very short rising time, shown in the inset. The dashed red curve is the fitting by using Eq. (12) below. In the absence of an optical trap, the dynamic equation of motion can be written as:

$$d\boldsymbol{p}(t)/dt = -\gamma_E \dot{\boldsymbol{x}}(t) + Q_{eff}\boldsymbol{E}_\infty \quad . \qquad (12)$$

For the left-hand side of Eq. (12), we sum up the momenta of the fluid and the solid at each finite element mesh within the inner/outer interface at each time step, and then evaluate its time derivative by using the data from successive time steps. The evaluation of the momenta is necessary since there are relative motions between the fluid and the solid. On the right hand side, the applied electric field is known, and for $\dot{\boldsymbol{x}}(t)$ we use the center of mass velocity of the evaluated unit. The two parameters $\gamma_E$ and $Q_{eff}$ are then varied as the fitting parameters to best reproduce the simulation data. For $a$=0.75 μm and $\lambda_D$ =96.1 nm, the best fit yields values of $\gamma_E/\gamma_S$ =1.7 and $|Q_{eff}|$=9.3×10$^{-16}$ C; they are in excellent agreement with the experimentally measured values of 1.69±0.27 and 9.1±1.4×10$^{-16}$ C, respectively. This indicates a consistent picture of drag coefficients evaluated from dynamic and static simulations, as well as consistency with the experimental measurements.

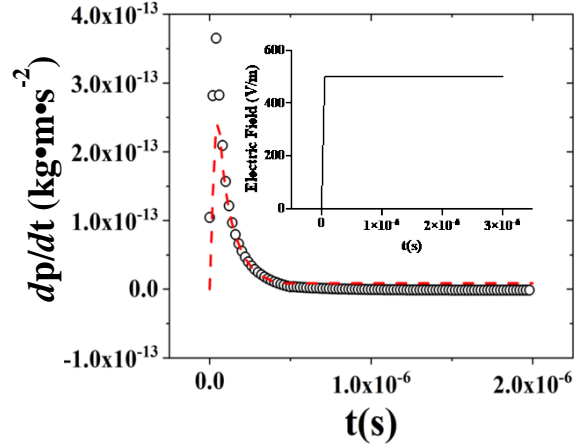

**Fig. 5.** Momentum time derivative evaluated inside the inner flow field (including the solid particle) plotted as a function of time. Black circles are from dynamic simulation and red dashed curve is fitted from Eq. (12). Inset shows the corresponding electric field as a function of time. It shows a linear increase within $5\times10^{-8}s$ to reach the saturation level.

The fact that the inner flow field is carried along by the charged particle also means that the interface that separates the inner and outer flow fields represents a slip surface/plane.

**Consistency between the framework of O'Brien and White, and Long et al, with the**

**present theory and experimental results**

We return to Eq. (1) and show that the inner flow field and its well-defined interface with the outer flow field constitute the missing piece in the theoretical framework of O'Brien and White[22], and Long et al.[21], that can tie together our theoretical predictions and the experimental results.

In a seminal paper, O'Brien and White simplified the governing equations for electrophoresis by assuming the coupled hydrodynamic and electric forces on a particle undergoing electrophoresis can be decoupled and solved separately[22]. The problem therefore becomes a superposition of two problems: the pure force problem (particle held fixed in a flow field with no applied field), and the pure electrophoresis problem (particle in an electric field in an electrolyte which is at rest at points far from the particle). Based on this analysis, Long et al. [21] showed that the velocity of the sphere is the sum of the two problems described above, based on global force balance[14,21]:

$$\boldsymbol{F}_{ext} - \gamma_s(\boldsymbol{v} - \mu_E \boldsymbol{E}_\infty) = 0, \tag{13a}$$

$$\boldsymbol{F}_{ext} + \boldsymbol{F}_{eletric} + \boldsymbol{F}_{visc} = 0. \tag{13b}$$

Here $\boldsymbol{F}_{electric}$ stands for the electric force, $\boldsymbol{F}_{visc}$ denotes the viscous drag force, and $\boldsymbol{F}_{ext}$ denotes the non-electric external force. It should be reminded that $v$ stands for the relative velocity between the particle and the far field fluid. Here Eq. (13a) is noted to be the same as Eq. (1).

Since we have demonstrated by using dynamic simulation that the inner flow field is carried along by the charged sphere, here we wish to use the inner/outer interface to evaluate all the relevant forces, so as to test Long et al's analysis while simultaneously also demonstrate the fact that $\gamma_E \neq \gamma_S$. The latter can be seen from Eq. (13a) that, if $\boldsymbol{F}_{ext} = 0$, then we must have $\boldsymbol{v} = \mu_E \boldsymbol{E}_\infty$. When that happens, the force balance, Eq. (13b), becomes $\boldsymbol{F}_{eletric} + \boldsymbol{F}_{visc} = 0$, i.e., a test of the electrophoretic drag also becomes possible since we can write $\boldsymbol{F}_{eletric} = Q_{eff}\boldsymbol{E}$ and $\boldsymbol{F}_{visc} = \gamma_E \boldsymbol{v}$, both evaluated at the inner/outer interface.

All simulations were performed with surface charge density $\sigma = -7 \times 10^3 e/\mu m^2$,

particle radius $a$=0.75 μm and ionic strength 0.01 mM ($\lambda_D$=100 nm).

To test Eqs. (13a) and (13b) at the interface that divides the inner and outer flow fields, $F_{electric}$ in Eq. (13b) is the sum of electric forces inside the interface that divides the inner and outer flow fields. This force can be expressed as:

$$F_{electric}\,|_{a+\Delta} = E_\infty Q_s + 2\pi \int_{-1}^{1} dcos\theta \int_{a}^{a+\Delta} r^2 dr \rho(r) \frac{Q_s \hat{r}}{\varepsilon_o \varepsilon_f r^2} + 2\pi \int_{-1}^{1} dcos\theta \int_{a}^{a+\Delta} r^2 dr \rho(r)(-\nabla\psi), \quad (14a)$$

where $\rho(r) = e[p(r) - n(r)]$. On the other hand, $F_{visc}$ stands for the drag force produced by the flow over the same inner/outer interface. It can be written as:

$$F_{visc}|_{a+\Delta} = 2\pi(a+\Delta)^2 \int_0^\pi (\sigma_{rr} cos\theta - \sigma_{r\theta} sin\theta) sin\theta\, d\theta, \quad (14b)$$

where $\sigma_{rr}$ and $\sigma_{r\theta}$ are the normal and tangential elements of the hydrodynamic stress tensor. From Eq. (13b), $F_{ext}$ is the negative of the sum of these two forces.

By maintaining the solid particle stationary and varying the boundary value of the far field fluid velocity $v$, we evaluate the magnitude of $F_{ext}$ as a function of $v$, with the externally applied electric field maintained at $E_\infty$=500V/m. We show in Fig. 6(a) that the external force $F_{ext}$ displays a linear dependence on the relative velocity between the solid particle and the bulk fluid, with a slope given by 1.03 $\gamma_S$. That is, Eq. (13a) is verified: $\frac{\Delta(F_{visc} + F_{elec})}{\Delta(\mu_E E_\infty - v)} = \gamma_S = \frac{F_{ext}}{\Delta v}$.

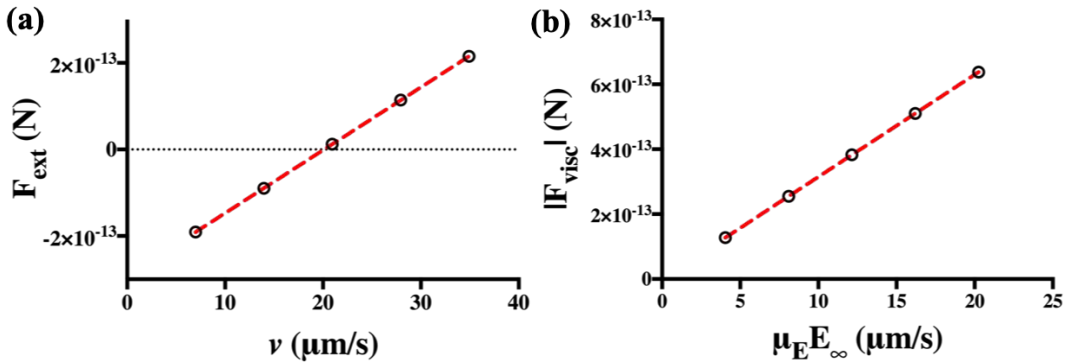

**Fig. 6 The difference between $\gamma_S$ and $\gamma_E$, using Eq. (13). (a)** External force $F_{ext}$ at interfacial region shows linear dependence with sphere velocity $v$. The open circles indicate five different values of $v$. The red dashed line denotes the best fitting with a slope 1.45×10$^{-8}$ kg·s$^{-1}$, which is very close to the Stokes drag coefficient 1.41×10$^{-8}$ kg·s$^{-1}$ (= $6\pi\eta a$, 3% lower than the slope). The external force becomes zero when $v$ is exactly at the electrophoretic velocity. The external electric field strength was maintained at $E_\infty$=500V/m. **(b)** Viscous force $F_{visc}$ at interfacial region with sphere velocity $v=\mu_E E_\infty$. The external electric field strength is varied from 100V/m to 500V/m in 100V/m steps. The red dashed line indicates fitting with slope $2.9\times10^{-8}$kg·s$^{-1}$(=2.1×6πη$a$). All simulations were performed with surface charge density $\sigma = -7\times 10^3 \, e/\mu m^2$, radius $a$=0.75 μm and ionic strength 0.01 mM ($\lambda_D$=100 nm).

It should be noted that even though the net $F_{ext}$ is evaluated at $r = a + \Delta$, the fact that the inner flow field is carried along by the solid particle implies the net force $F_{ext}$ to be the same as that evaluated at the center of mass of the particle, as well as at $r = a$.

In Fig. 6(b), we focus on the case when $\boldsymbol{F}_{ext} = 0$ so that $\boldsymbol{F}_{eletric} = Q_{eff}\boldsymbol{E} = \boldsymbol{F}_{visc} = \gamma_E \boldsymbol{v}$. By evaluating the viscous drag at the inner/outer interface and varying the external electric field strength from 100V/m to 500V/m in 100V/m steps, we plot the resulting $\boldsymbol{F}_{visc}$ as a function of the electrophoretic velocity $\mu_E \boldsymbol{E}_\infty$. The slope gives a value that is 2.1 times the Stokes drag coefficient, which is somewhat higher than the experimentally measured value of $\gamma_E/\gamma_S = 1.7 \pm 0.3$. However, both show unmistakable deviation from the Stokes drag coefficient. Therefore, we have explicitly used Long et al's analysis, based on the work of O'Brien and White, to show both the correctness of the equation $\boldsymbol{F}_{ext} - \gamma_s \Delta \boldsymbol{v} = 0$, as well as the necessary difference between the Stokes drag coefficient and the electrophoretic drag coefficient, *both evaluated at the inner/outer interface*.

**CONCLUSIONS**

We have measured electrophoresis drag coefficient $\gamma_E$ by using an AC electric field to drive a charged particle in an optical trap. As far as we are aware, this is the first time such non-Stokes electrophoretic drag coefficients are determined experimentally. Our measurements were possible because our experimental arrangement permitted a time-varying electrophoretic motion of the charged particle with a speed directly proportional to that of the corresponding electric field strength. The non-electric, out-of-phase force from optical trap, fully counter-balanced by the out-of-phase electric force on the particle, does not play a role in hindering the electrophoretic motion. From both the experimental measurements and the direct numerical solutions, we show that the observed non-Stokes electrophoretic drag coefficient can be quantitatively described by the uniquely distinctive inner/outer interface as the reference plane for the evaluation of the hydrodynamic drag. The measured effective charge, $Q_{eff}$, is consistent with the measured drag coefficient as required by force balance at the inner/outer interface flow fields which is at a distance a few $\lambda_D$'s away from the solid surface. The present experimental-theory study provides new, microscopic insights into the classic problem of electrophoresis. Such understanding can shed light on new applications of electrophoresis to investigate biological nanoparticles. More broadly, the experimental approach taken by this study and the microscopic view of an interface between the inner-outer flows might shed light to the outstanding problems of hydrodynamic forces and drags of other types of phoretic particles or swimming microorganisms[34].

**MATERIALS AND METHODS**

An optical trap was effected by an IR (wavelength=1064 nm) laser coupled into an oil-immersion objective lens (100X, NA=1.3, Olympus). A second IR laser beam (wavelength=980 nm), aligned and focused by the same objective lens to be par-focal with the trapping laser focus, was used for particle tracking. A schematic diagram of the measurement apparatus is shown in Fig. S1. Here, we use the optical trap both as a tool to control/monitor the position of the particle as well as a calibrated force sensor[35-37] with a stiffness constant $k_{trap}$=17.9 ± 0.1 pN/μm. Experimental details are given in Section S3 of SM.

Polystyrene (PS, Thermo Scientific, catalog #5153A) particles with a mean radius $a$=0.75 μm were dispersed at volume fractions below 0.0001 in solutions of varying concentrations of potassium chloride electrolyte with a Debye length $\lambda_D$ ranging from 96 to 4.3 nm (KCl concentration 0.01 to 5mM, κ$a$ ($\equiv a/\lambda_D$) ranging from 7.8 to 174). To reduce the effect of dissolution of $CO_2$ in DI water, all electrolytes were prepared with fresh DI water filtered with resin. The colloidal solutions were inserted in a glass chamber with vertical height ~200 μm. To avoid the effects of the Faxén drag on the particle[38-40] and the flow due to any surface electroosmosis flow near the glass surface[41], we kept the colloid sphere 8-15 μm above the lower glass plate.

The electrophoretic chambers were fabricated by coating a glass microscope slide substrate with Sigmacote (Sigma-Aldrich), a solution of chlorinated organopolysiloxane in heptane that readily forms a covalent, microscopically thin film on glass, with a purpose to suppress surface charge and to minimize surface electroosmotic flow. Subsequently gold wires for applying the electric field were placed on the glass substrate with a separation of ~1 cm. An aqueous solution of dilute particles was then dipped onto the substrate, and a Sigmacote-coated cover-glass was used to seal the chamber with wax sealant. External electrical wires were welded with the gold wires. Electric field was measured by inserting a second pair of parallel gold wires, separated by 5 mm, into the sample to obtain the voltage drop between them. The optical path diagram is shown schematically in Fig. S1 in SM. The electrophoresis chamber containing the particles was mounted on an inverted microscope (Olympus IX81). To minimize the contribution to the optical trapping effect, the 980 nm tracking beam power was two orders of magnitude lower than that of the 1064 nm laser. Movements of the particle, tracked by the 980 nm laser beam, were detected by a quadrant photodiode (QPD, S7479, Hamamatsu). The voltage reading of the QPD was maintained to be within the linear range of the particle displacement from the trapping center. A lock-in amplifier (SR830, Stanford Research) was used to record the phase of particle motion relative to that of the sinusoidal voltage applied to PZT to form an oscillatory optical tweezers. The wide-field images of the particle were captured by a CCD camera for the purpose of optical alignment.

**ACKNOWLEDGEMENTS**

PS wishes to acknowledge the funding support by Hong Kong RGC grants GRF16303415and C6004-14G for this work. HDO wishes to acknowledge the support by US NSF grant 0923299 and the Emulsion Polymers Institute. PS, ML, and SX wish to thank Chun Liu for advice regarding the necessary condition(s) for the appearance of hydrodynamic vortices. HDO and MTW would like to thank Kathryn Reddy, a former undergraduate student of Fordham University who participated in the initial experiments of this study during her REU at Lehigh.

HDO would also like to thank Prof. Joel A. Cohen, Prof. Daan Frenkel and Dr. Alois Würger for their critical comments and helpful suggestions.

AUTHOR CONTRIBUTIONS

PS and HDO initiated and supervised the research. ML and MTW carried out the experiments. ML and PS contributed to theory and simulations with help from SX. ML. HDO and PS analyzed the data with help from MTW. ML, PS and HDO wrote the draft manuscript. All participated in revising the manuscript to its final form. ML and MTW contributed equally to this work.

# SUPPLEMENTARY MATERIALS

## to

## Non-Stokes Drag Coefficient in Single-Particle Electrophoresis: New Insights on a Classical Problem

*Maijia Liao* (廖麦嘉), *Ming-Tzo Wei*（魏名佐）, *Shixin Xu*（徐士鑫）,
*H. Daniel Ou-Yang*（歐陽新喬）, *Ping Sheng*（沈平）

S1   EXPERIMENTAL SETUP
S2   THE SMOLUCHOWSKI SOLUTION
S3   CALIBRATION OF OPTICAL TWEEZERS
S4   COMPARISON BETWEEN MEASURED AND LITERATURE $\mu_E / \zeta$ RELATIONS
S5   SIMULATION DETAILS
S6   NON-INERTIAL GENERATION OF VORTICITY INSIDE THE DEBYE LAYER

## S1: EXPERIMENTAL SETUP

In Fig. S1 below, we give a schematic diagram of the experimental setup used to measure the electrophoretic drag coefficient.

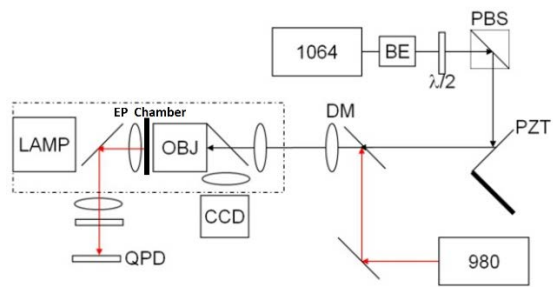

**Fig. S1. A schematic diagram of the experimental setup.** BE: beam expander; λ/2: half-wave plate; PBS: polarizing beam splitter; PZT mirror: mirror mounted on a PZT piezo-electric transducer; DM: dichromatic mirror; OBJ: microscopic objective; EP Chamber: electrophoresis chamber; QPD: quadrant photodiode.

## S2: THE SMOLUCHOWSKI SOLUTION

Here we first derive an expression relating the fluid velocity to the zeta potential in a cylindrical channel with charged boundary, to be followed by the Smoluchowski solution to the electrophoresis flow field. In the first part, we show that the Smoluchowski approach is to use the appearance of the Laplacian operator and the net charge in both the Poisson equation and the Navier-Stokes equation, for relating the fluid velocity to the electrical potential without actually having to solve any differential equations.

Consider a cylindrical channel with uniformly charged wall. The counter ions naturally form a screening Debye layer. Since there is net charge $\rho$ in the Debye layer, the electrical potential $\psi$ in the fluid is governed by the Poisson equation:

$$\frac{\partial^2 \psi}{\partial r^2} + \frac{1}{r}\frac{\partial \psi}{\partial r} = -\frac{\rho}{\varepsilon}, \tag{S1}$$

where $r$ denotes the radial coordinate and $\varepsilon$ denotes the dielectric constant. It is to be noted that the electrical potential variation and the consequent electric field in the Debye layer are both along the radial direction.

If an electric field $E_z$ is applied along the axial direction $z$, the field can exert a force on the net charge density present in the Debye layer. This force density would appear on the right-hand side of the Navier-Stokes (NS) equation to drive the flow. In steady state, the NS equation can be written as:

$$\eta\left(\frac{\partial^2 u_z}{\partial r^2} + \frac{1}{r}\frac{\partial u_z}{\partial r}\right) = \frac{\partial P}{\partial z} - \rho E_z, \tag{S2}$$

where $\eta$ is the fluid viscosity and $P$ the pressure. By equating $\rho$ that appears in both Eq. (S1) and Eq. (S2), we obtain

$$\frac{\partial^2 u_z}{\partial r^2} + \frac{1}{r}\frac{\partial u_z}{\partial r} = \frac{1}{\eta}\frac{\partial P}{\partial z} + \frac{\varepsilon E_z}{\eta}\left[\frac{\partial^2 \psi}{\partial r^2} + \frac{1}{r}\frac{\partial \psi}{\partial r}\right]. \tag{S3}$$

Equation (S3) suggests that the fluid velocity $u_z$ and the electrical potential $\psi$ must be linearly related to each other, i.e. $u_z(r) = C_1 + C_2 \psi(r)$; with $C_1$ and $C_2$ to be determined by the boundary condition that $\psi = \zeta$ on the solid boundary plus the requirement of satisfying Eq. (S3). Since it is well known that in the presence of a constant pressure gradient the flow field in a cylindrical pipe must have a parabolic velocity profile, one can readily write down the relationship between the fluid velocity and the electrical potential as:

$$u_z = -\frac{\varepsilon E_z}{\eta}\left[\zeta - \psi\right] + \frac{a^2 - r^2}{4\eta}\left(-\frac{dP}{dz}\right). \tag{S4}$$

Here $a$ is the channel radius. For $dP/dz = 0$ and a large $a$ so that the electrical potential must be zero at the center of the channel, we have

$$u_z = -\frac{\varepsilon E_z}{\eta}\zeta = -\frac{\varepsilon\zeta}{\eta}E_z = \mu_E E_z. \tag{S5}$$

This is the well-known Smoluchowski relation for the electrophoresis mobility $\mu_E = -\varepsilon\zeta/\eta$. It relates the far-way fluid velocity to the stationary solid wall. When interpreted in the context of electrophoresis, the solid wall is the surface of a charged spherical particle moving relative to stationary fluid far away. Equation (S5) then expresses the linear relationship between the electrophoretic velocity and the applied electric field.

The relevant physics encoded by the Smoluchowski relation can be simply described in terms of Newton's laws. Since the electric field is acting on an overall electrically neutral system that comprises the surface charge and the screening counter ions, the center of mass momentum of the whole system should be zero. However, since the net charge in the Debye layer can be driven by the applied electric field so as to cause electroosmotic flow, it follows that the solid particle must move in the opposite direction in order to maintain the zero center of mass momentum. That is the observed electrophoretic velocity.

To express the above physics mathematically, Smoluchowski first observed that for a weakly charged surface and $\lambda_D \ll a$, the sphere plus its boundary layer have the appearance in the far field region of a neutral, non-conducting sphere[1,2]. The Laplace equation describing the outer region potential can be written as $\nabla^2\psi = 0$, with boundary conditions on the surface of the neutral sphere given by $\hat{r}\cdot\nabla\psi|_{r=a} = 0$, while at infinity it is given by $-\nabla\psi = \boldsymbol{E}_\infty$. The tangential field boundary condition at $r=a$ follows from the Smoluchowski argument[3] in which the essential assumption is the existence of a concentric spherical surface where the external electric field is aligned with the velocity field as required by the electroosmotic flow relation (S5). That is, the external electric field would act on the charged fluid inside the (ultrathin) Debye layer through the body force density term in the Stokes equation to induce a surface-tangential electroosmotic flow.

Since the applied uniform electric field $\boldsymbol{E}_\infty$ represents an electrical potential source with dipolar symmetry, the solution of the Laplace equation for the electric potential must consist (besides the source term) of a term that has the same dipolar symmetry. As a result, the electric field outside the sphere can be written as $\boldsymbol{E}(\boldsymbol{r}) = [\vec{I} + \frac{1}{2}\left(\frac{a}{r}\right)^3(\vec{I} - 3\hat{r}\hat{r})]\cdot\boldsymbol{E}_\infty$, where $a$ is used under the assumption of an ultrathin Debye layer. The second term in the square bracket above is precisely the electric field of a dipole with a coefficient given by $-a^3\boldsymbol{E}_\infty/2$, obtained by using the tangential electric field boundary condition on the surface of the neutral sphere as given above. An illustration of the electric field lines is shown schematically in Fig. S2.

From the Smoluchowski argument, the fluid velocity field in the particle's rest frame can be written as $\boldsymbol{u} = -\dfrac{\varepsilon\zeta}{\eta}\boldsymbol{E}_\infty - \dfrac{\varepsilon\zeta}{2\eta}\left(\dfrac{a}{r}\right)^3 (\boldsymbol{\ddot{I}} - 3\hat{\boldsymbol{r}}\hat{\boldsymbol{r}})\cdot\boldsymbol{E}_\infty$. The velocity component along the electric field direction can thus be expressed as:

$$\boldsymbol{u}(\boldsymbol{r})\cdot\hat{\boldsymbol{E}}_\infty = -\dfrac{\varepsilon\zeta}{\eta}\hat{\boldsymbol{E}}_\infty\cdot[\boldsymbol{\ddot{I}} + \dfrac{1}{2}\left(\dfrac{a}{r}\right)^3 (\boldsymbol{\ddot{I}} - 3\hat{\boldsymbol{r}}\hat{\boldsymbol{r}})]\cdot\boldsymbol{E}_\infty = -\dfrac{\varepsilon\zeta}{\eta}\left[1 + \dfrac{1}{2}\left(\dfrac{a}{r}\right)^3 (1 - 3\cos^2\theta)\right] E_\infty, \quad (S6)$$

From Eq. (S6), which expresses the fluid velocity in the rest frame of the electrophoretic particle, it follows that in the laboratory frame, the fluid velocity in the far field must have the functional form shown in Eq. (11b), i.e., $\boldsymbol{u}(\boldsymbol{r})\cdot\hat{\boldsymbol{E}}_\infty|_{lab} \sim P_2(\cos\theta)/r^3$, where $P_2(\cos\theta)$ denotes the second order Legendre polynomial. It should be noted that the $1/r^3$ far field behavior has been derived by using a much more rigorous mathematical approach.

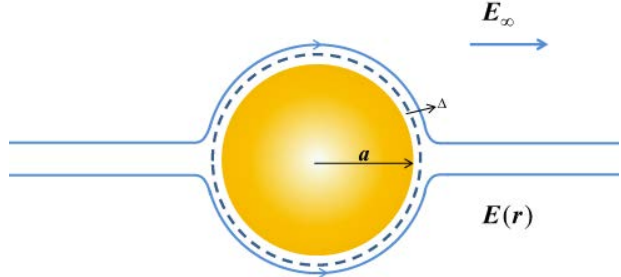

**FIG. S2** Illustration of the electric field in the outside region. Here *a* stands for the particle radius. The dashed line roughly indicates the position of the Debye layer. Outside electric field is denoted as $\boldsymbol{E}(\boldsymbol{r})$, and the electric field at infinity is denoted as. The electric field lines bend tangentially at the interfacial boundary as constructed by the Smoluchowski solution.

## S3: CALIBRATION OF OPTICAL TWEEZERS

Optical tweezers are known to provide a harmonic potential for the trapped dielectric particle near the center of the trap. To characterize the force constant $k_{trap}$, we study the motion of a trapped particle held by an oscillatory tweezer in a viscous medium. The equation of motion can be expressed as:

$$m\ddot{x}(t) = -\gamma_S \dot{x}(t) + k_{trap}(Ae^{-i\omega t} - x(t)), \quad (S7)$$

where $m$ is the mass of the particle, $x$ is the displacement of a particle from the tracking beam center, $A$ is the amplitude of the oscillatory laser trapping beam with angular frequency $\omega$, and $\gamma_S$ stands for the Stokes drag coefficient (=$6\pi\eta a$, with $a$ being the particle radius and $\eta = 1\times 10^{-3}$ Pa.s is the viscosity of water). By taking into account the small mass and low acceleration of the particle, one can neglect the term on the left-hand side of Eq. (S8). This leads to a simplified equation:

$$\gamma_S \dot{x}(t) + k_{trap} x(t) = k_{trap} A e^{-i\omega t}, \tag{S8}$$

which has a solution given by

$$x(t) = D(\omega) e^{-i[\omega t - \delta(\omega)]}, \tag{S9}$$

with the displacement amplitude $D(\omega)$ expressible as

$$D(\omega) = \frac{A k_{trap}}{\sqrt{k_{trap}^2 + (\gamma_S \omega)^2}}, \tag{S10}$$

and

$$\delta(\omega) = \tan^{-1} \frac{\gamma_S \omega}{k_{trap}}. \tag{S11}$$

Here $\delta(\omega)$ stands for the phase of the particle motion relative to that of the oscillatory optical trap. In Eqs. (S10) and (S11), $k_{trap}$ is the only fitting parameter, owing to the fact that the values of $D(\omega)$ and $\delta(\omega)$ can be experimentally determined and other parameters are known quantities. For a PS sphere with $a$=0.75 μm, a value of $k_{trap}$=17.8±0.3 pN/μm and 18.0±1.1 pN/μm can yield good agreement between the measured and calculated values of $D(\omega)$ and $\delta(\omega)$, respectively, as shown in Fig. S3.

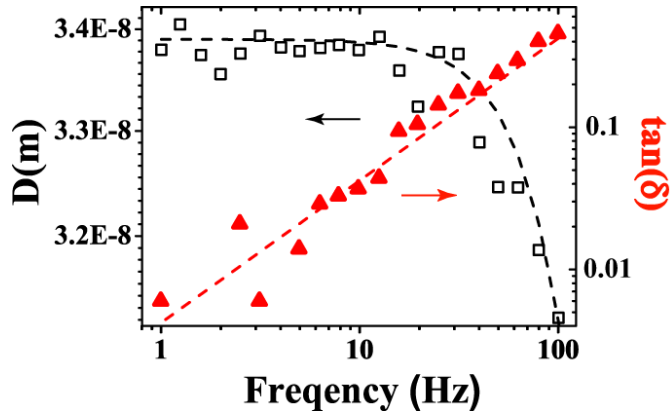

**FIG. S3.** Experimentally measured displacement and phase delay. The measured amplitude and relative phase for a 0.75 μm radius polystyrene particle held by an oscillatory optical tweezer in de-ionized (DI) water, plotted as a function of the oscillation frequency. The symbols represent the experimental data and the dashed lines are the fits with the spring constant $k_{trap}$ as the only fitting parameter in Eq. (S10) and (S11). The best fit of amplitude gives $k_{trap}$ =17.8±0.3 pN/μm and the best fit of phase data gives $k_{trap}$ =18.0±1.1 pN/μm.

To ensure the use of Stokes drag for calibration of $k_{trap}$ to be correct, we have also determined $k_{trap}$ by trapping the same particle in a stationary trap. By using the equal-partition theorem we obtain

$$k_{trap} = \frac{2 k_B T}{<x^2>}, \tag{S12}$$

where $k_B$ is the Boltzmann constant, $T$ the absolute temperature, and $<x^2>$ the mean-square Brownian displacement of the particle in the trap. The value of $k_{\text{trap}}$ so determined agreed with that obtained from Eqs. (S11) and (S12), ensuring the correct assumption that the drag coefficient in Eq. (S8) was indeed Stokes.

## S4: COMPARISON BETWEEN MEASURED AND LITERATURE $\mu_E / \zeta$ RELATIONS

Since mobility $\mu_E$ constitutes one of the most measured quantities in literature, our AC / optical tweezer measured values, which were verified to be the same as by in-situ DC experiments on the same particle, should be compared to the literature values. However, as shown in Section S2, the mobility is intimately connected to the surface charge density, which in turn is directly related to the zeta potential, hence in comparing with the literature values of the mobility one must specify the zeta potential as well. Also, in our simulations the measured mobility value is treated as the input boundary condition to obtain the correct value of the surface charge density that can yield force balance. Since surface charge density and zeta potential correspond to each other in a one-to-one fashion, it follows that the comparison of the simulated zeta potential values (with the experimental mobility the only input) with the literature values at the same mobility constitutes a definitive verification.

In Fig. S4(a), we summarize the values of $\mu_E$ obtained under different ionic strengths. Here $\kappa = 1/\lambda_D$ is proportional to the square root of the ionic density in the liquid. It is seen that there is a general trend of decreasing mobility with increasing salt concentration. To compare our measured values with the theory prediction from H. Ohshima et. al[4], for each value of the mobility at a given value of the salt concentration, one can obtain a value for the zeta potential which is denoted by the open symbol in Fig. S4(b). The solid symbols in Fig. S4(b) represent the zeta potential values obtained from our simulations (with the experimental values of the mobility as input boundary condition). Very good agreement is seen.

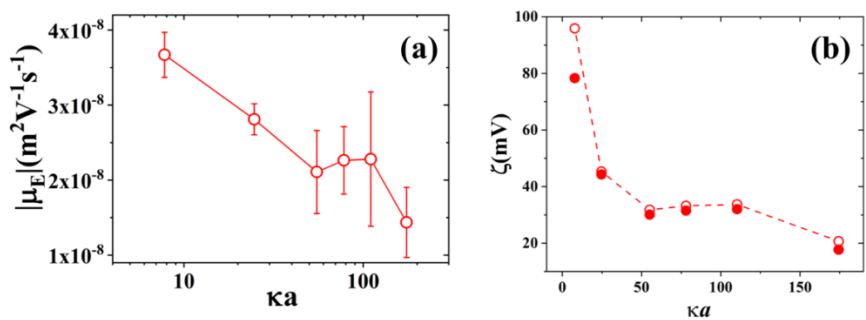

**FIG. S4. (a)** Measured mobility plotted as a function of κ$a$. Error bars indicate the errors in amplitude and phase measurements.
**(b)** The open symbols linked by dashed lines denote the values of the zeta potential determined from Ohshima's analytical electrophoresis expression (*4*) for given values of κ$a$ and mobility. Solid symbols indicate the simulated zeta potential values obtained by using the measured mobility in (a) to constrain the velocity $v=\mu_E E_\infty$ of the solid particle relative to the far field. Good agreement is seen.

## S5: SIMULATION DETAILS

We obtained converging numerical solutions to the nonlinear Nernst-Planck-Poisson equations, coupled to the Stokes equation in spherical coordinate with azimuthal symmetry (with the axis of symmetry being along electric field direction), by using the COMSOL Multiphysics 4.4 package. The electrostatic module, transport of diluted species module, and the creeping flow module were used in combination for the solution of the static problem. We adjusted the surface charge density from low to high by using the continuation method in order to improve the convergence of the non-linear solutions. Here the continuation method denotes the use of previous iteration's solution to be the initial guess of the subsequent iteration with slightly altered parameter value(s). The boundary conditions for Eqs. (10a)-(10i) are given in the main text. The spherical domain size of the simulation is 90 μm in its radius, 120 times of the particle radius 0.75 μm. The mesh size is 0.62 nm near the particle surface. At liquid-solid interface, 30 boundary layers with a stretching factor 1.1 were generated to match the mesh size of 0.62 nm near the particle surface. This is to ensure a smooth transition of the mesh size at the interfacial region.

In dynamic simulations, a deformed mesh interface is used. Arbitrary Lagrangian-Eulerian module is used when performing time dependent simulation using Eqs. (10a)-(10i). Surface charge obtained from static simulation through measured mobility is used as boundary condition on particle surface. In order to substantiate the proposition that the measured hydrodynamic drag coefficient should be evaluated by using the inner/outer interface (dashed white curve in Fig. 3(b)) as the reference plane, we have carried out dynamic simulations by using moving mesh. We applied a time-varying electric field that increases linearly from zero to its saturation level within 0.05μs. Dynamic simulation is performed in the laboratory frame. To obtain time evolution of velocity, Eq. (10i) is updated at a time step of $5\times10^{-4}$μs.

## S6: NON-INERTIAL GENERATION OF VORTICITY INSIDE THE DEBYE LAYER

For the appearance of vortices in the solution of NS equation there must be a sufficiently large source for the vorticity, defined as $\boldsymbol{\omega} = \nabla \times \boldsymbol{u}$. For the viscous incompressible flow with electrical body force, the Navier-Stokes equation is given by

$$\frac{\partial \boldsymbol{u}}{\partial t} + \boldsymbol{u}\cdot\nabla\boldsymbol{u} = -\frac{1}{\rho}\nabla P + \frac{\eta}{\rho}\nabla^2\boldsymbol{u} + \frac{1}{\rho}\boldsymbol{f} \ , \tag{S13}$$

where $\boldsymbol{f} = -e(p-n)\nabla\psi$ denotes the electrical body force density. By taking the curl of both sides of the NS equation, the steady state dimensionless vorticity equation with electrical body force can be written as:

$$\left(\nabla^2 \boldsymbol{\omega}^*\right) + \text{Re}\left(\boldsymbol{\omega}^* \cdot \nabla\right) \boldsymbol{u}^* + \frac{\text{Re}\,\varepsilon \zeta^2}{d^2 \mu_E^2 E_\infty^2 \rho}\left[\nabla\left(\nabla^2 \psi^*\right)\right] \times \nabla \psi^* = 0, \quad (S14)$$

where $\boldsymbol{u}^* = \dfrac{\boldsymbol{u}}{v}$, $l^* = \dfrac{l}{d}$, $t^* = \dfrac{t}{d/v}$, $\boldsymbol{\omega}^* = \dfrac{\boldsymbol{\omega} d}{v}$, $\psi^* = \dfrac{\psi}{\zeta}$, and Re= $\rho u d / \eta$ is the dimensionless Reynolds number. Here $d$ denotes the sphere diameter and $v$ denotes the sphere velocity relative to the far field bulk fluid, and $\zeta$ is the zeta potential. When there is no electrical body force, it is generally accepted that vortices appear when the Reynolds number is large (Re > 10)[5], so that the inertial term $\text{Re}\left(\boldsymbol{\omega}^* \cdot \nabla\right) \boldsymbol{u}^*$ represents a large source for the vorticity Poisson equation. In the present case, the second term in Eq. (S14) is negligible, since the Reynolds number in our experimental systems is ~$10^{-6}$. At the same time, since $p$ and $n$ are spatially varying near the particle's surface, hence the curl of their product with $\nabla \psi$ does not vanish in such regions. Therefore, we essentially have a Poisson equation for $\boldsymbol{\omega}^*$ with a substantial source term that arises from the electrical body force density near the particle's surface. In the current work Re~$10^{-6}$, $\zeta \sim 25mV$, $d \sim 1\mu m$, $\mu_E \sim 10^{-8} m^2/(V \cdot s)$, $E_\infty = 500 V/m$, we have the dimensionless number $\dfrac{\text{Re}\,\varepsilon \zeta^2}{d^2 \mu_E^2 E_\infty^2 \rho}$ on the order of 10. To illustrate the spatial distribution of this source term, we plot the scalar quantity $\hat{\boldsymbol{\varphi}} \cdot \left[\nabla\left(\nabla^2 \psi^*\right) \times \nabla \psi^*\right]$ near the particle surface as shown in Fig. S5. It is seen that close to the particle surface, the magnitude of $\hat{\boldsymbol{\varphi}} \cdot \left[\nabla\left(\nabla^2 \psi^*\right) \times \nabla \psi^*\right]$ can reach ~$10^3$. Here the unit vector $\hat{\boldsymbol{\varphi}}$ is along the azimuthal angular direction. Hence in Eq. (S14) the third term can be a significant vorticity source, and it is in exactly the same region where we observed the ring of vortices.

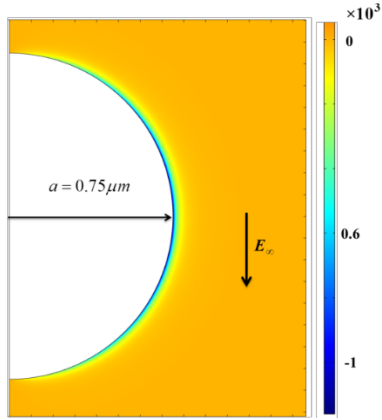

**FIG. S5.** Non-zero vorticity source term near the particle surface. The distribution of $\hat{\boldsymbol{\varphi}} \cdot \left[\nabla\left(\nabla^2 \psi^*\right) \times \nabla \psi^*\right]$ near the particle surface is plotted by using the color scheme where yellow is close to zero. Simulation is performed with particle radius $a$=0.75 μm and surface charge density σ=5660e/μm$^2$ under an electric field strength of $E_\infty$=500 V/m and ionic strength 0.01mM ($\lambda_D$=96.1 nm).

However, it should be noted that non-zero vorticity by itself does not guarantee the appearance of vortices. For example, the vorticity may be nonzero even when all the stream lines are straight and parallel. The appearance of vortices requires the reversal in the direction of fluid flow. This is exactly what happens near the surface of the solid particle, where the large local net charge concentration means strong electroosmotic flow along the field direction near the equatorial plane, where the electric field is tangential to the surface. However, in the lab frame, close to the solid particle's surface the flow field must reverse its direction (owing to the nonslip boundary condition) so as to be aligned with the solid particle's velocity (noted to be opposite to the field direction). Thus, the nonzero vorticity, plus the strong local electroosmotic flow that results in velocity reversal, provide the necessary condition for the generation of vortices in electrophoretic dynamics in the laboratory frame.